\documentclass[twocolumn,english,prl]{revtex4-1}
\usepackage[T1]{fontenc}
\usepackage[latin9]{inputenc}
\usepackage{graphicx}
\usepackage{babel}
\begin{document}

\title{Elongation and percolation of defect motifs in anisotropic packing
problems}

\author{Zhaoyu Xie}

\affiliation{Department of Physics \& Astronomy, Tufts University, 574 Boston
Ave, Medford MA 02155}

\author{Timothy J. Atherton}

\affiliation{Department of Physics \& Astronomy, Tufts University, 574 Boston
Ave, Medford MA 02155}
\begin{abstract}
We examine the regime between crystalline and amorphous packings of
anisotropic objects on surfaces of different genus by continuously
varying their size distribution or shape from monodispersed spheres
to bidispersed mixtures or monodispersed ellipsoidal particles; we
also consider an anisotropic variant of the Thomson problem with a
mixture of charges. With increasing anisotropy, we first observe the
disruption of translational order with an intermediate orientationally
ordered hexatic phase as proposed by Nelson, Rubinstein and Spaepen,
and then a transition to amorphous state. By analyzing the structure
of the disclination motifs induced, we show that the hexatic-amorphous
transition is caused by the growth and connection of disclination
grain boundaries, suggesting this transition lies in the percolation
universality class in the scenarios considered.
\end{abstract}
\maketitle

\section{Introduction}

\begin{figure*}
\begin{centering}
\includegraphics[width=1\textwidth]{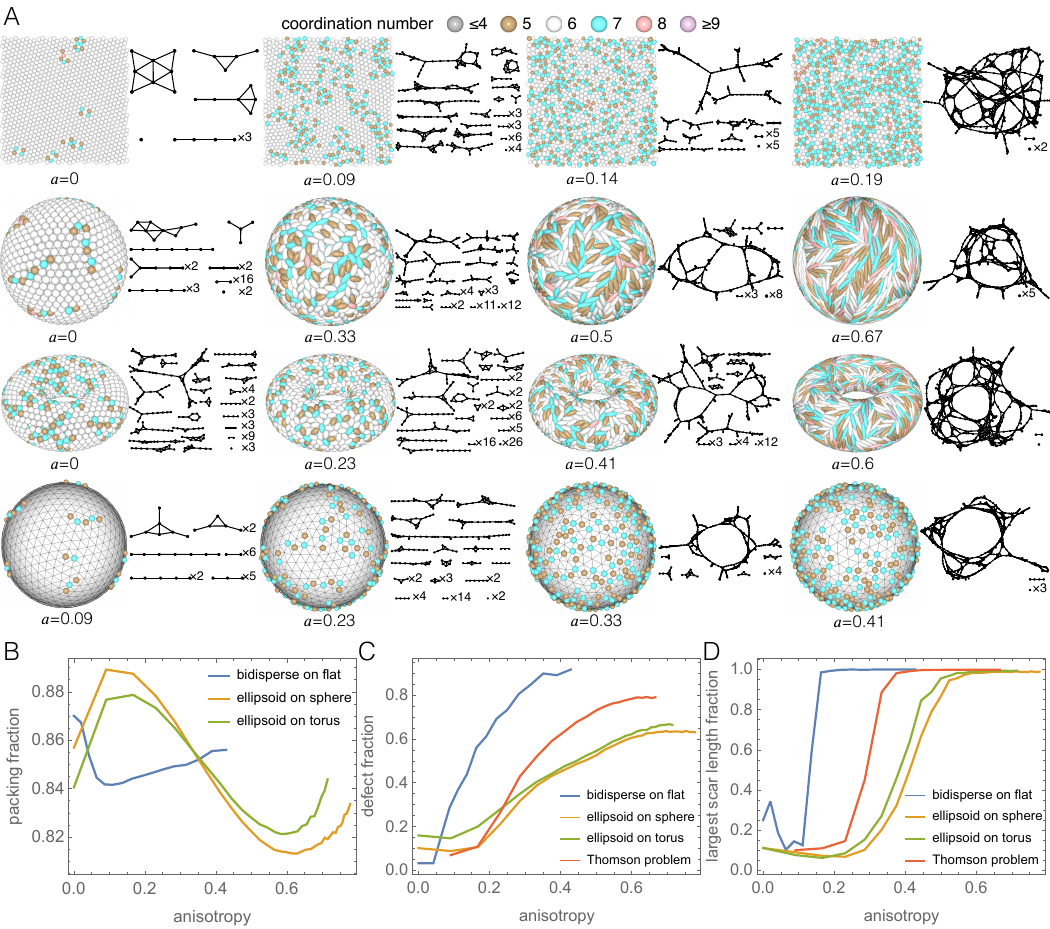}
\par\end{centering}
\caption{\label{fig:Packings-and-associated}\textbf{Elongation and percolation
of defect networks occurs with increasing anisotropy in a variety
of anisotropic packing problems.} (A) Packings as a function of anisotropy
with particles colored by coordination number; the defect subgraph
for each packing is calculated from the neighbor graph by retaining
only non-hexagonally coordinated vertices. (B) Packing fraction as
a function of anisotropy. (C) Fraction of defects and (D) the fraction
of largest scar length as a function of anisotropy show a characteristic
shape that is a signature of the percolation transition. }
\end{figure*}

Packing problems, where a set of objects are arranged in a specified
container to optimize the density, are an important model of many
materials including granular media, colloids and amorphous solids\citep{bowick2009two,torquato2010jammed,baule2014fundamental,manoharan2015colloidal,li2016assembly,torquato2018perspective}.
In two-dimensional unbounded Euclidean space, the highest-density
packing of disks is the hexagonal lattice where each particle is surrounded
by six neighbors. Numerous situations where this highly regular crystalline
arrangement becomes disordered have been explored: If the boundaries
of the container are not commensurate with the lattice\citep{nelson1982order,rubinstein1982order},
or if the packing occurs on a curved surface so that the lattice is
incompatible with the curvature\citep{bausch2003grain,bowick2000interacting,irvine2010pleats,jimenez2016curvature},
or if the particles are no longer circular and equal in size\citep{williams2003random,donev2004improving,chaikin2006some,yao2014polydispersity},
the overall arrangement may lose either translational or orientational
order, or both.

Topological defects, deviations from crystalline order that cannot
be removed by continuous deformations, are an invaluable concept to
understand the resulting packings. The elementary defect in a hexagonal
lattice is a \emph{disclination}, a site that possesses a coordination
number $n\neq6$; these tend to disrupt the orientational order as
they promote rotation of the lattice vectors. Interactions between
disclinations are analogous to electrostatics, motivating the definition
of a topological charge $q=(6-n)$. Other defect motifs that occur
include \emph{dislocations}, disclination dipoles, \emph{scars}, chains
of disclinations of alternating sign that are induced on curved surfaces
to accommodate the curvature\citep{bowick2000interacting}, \emph{pleats}
that are bound to the edge of an open curved manifold\citep{irvine2010pleats}
and \emph{grain boundaries} that separate uncorrelated regions of
crystalline order\citep{nelson2002defects}.

Consider perturbing the crystalline packing of monodispersed disks
of radius $r$ by replacing some fraction $\chi$ with a larger radius
$R$ such that $R/r>1$. We may define a dimensionless parameter,
the bidispersity $b=\left(R-r\right)/(R+r)\in[0,1]$, to describe
the deviation from monodispersity. As $b$ increases, Nelson, Rubinstein
and Spaepen (NRS)\citep{rubinstein1982order,nelson1982order} predict
the following sequence: first dislocations appear introducing stacking
faults that disrupt long range translational order. There then exists
an intermediate \emph{hexatic} phase that possesses either long range
or power-law orientational order as the lattice vectors of adjacent
patches of crystal remain correlated. Further increasing $b$ leads
to an amorphous phase that lacks both translational and orientational
order. The hexatic phase is a zero-temperature analog of the intermediate
hexatic phase that mediates melting in the Kosterlitz-Thouless-Halperin-Nelson-Young
theory\citep{halperin1978theory,nelson1979dislocation,young1979melting}.
This phase transition into the amorphous phase triggered by the unbinding
of topological defects also occurs in systems of hard disks\citep{bernard2011two,qi2014two},
hard regular polygons\citep{anderson2017shape}, soft disks with repulsive
power-law interactions\citep{kapfer2015two,hajibabaei2019first} and
active Brownian particles\citep{digregorio2018full,paliwal2020role}.

On a curved surface, such as a sphere, the NRS picture must be modified
because defects are required even in the ground state, leading to
a regime referred to as \emph{spherical crystallography}\citep{bowick2000interacting,bausch2003grain,bowick2009two},
and vector transport properties of the curved surface complicates
the measurement of long-range orientational correlations\citep{giarritta1992statistical,guerra2018freezing,vest2018glassy}.
Isolated disclinations occur for a small number of particles $N$
while for large $N$ these become spatially extended \emph{scars}
trading off the free energy cost of creating additional defects in
order to reduce deformation of the lattice\citep{bausch2003grain,bowick2000interacting}.
On spheres, these structures are icosahedrally ordered\citep{guerra2018freezing},
while the distribution for less symmetric surfaces is driven by the
distribution of Gaussian curvature\citep{seung1988defects,bowick2000interacting,vitelli2006crystallography,giomi2007crystalline,giomi2008defective,bowick2009two,irvine2010pleats}.

We recently showed for packing bidispersed spheres on a sphere that
as the bidispersity is increased from zero, the defect motifs begin
to elongate above a critical value of bidispersity $b=0.08$, continue
to grow and eventually form a connected structure; at the same time
the orientational order parameter becomes increasingly short range\citep{mascioli2017defect}.
 The hexatic-amorphous transition in this specific system may therefore
be equivalently viewed as elongation and percolation of the scars,
providing a connection between the regimes of spherical crystallography
and random close packing and, by leveraging the results of percolation
theory\citep{grimmett1999percolation,stauffer2018introduction}, successfully
predicting the distribution and microstructure of the defects.

An obvious question arises whether the percolation of defects also
applies to the NRS picture since the behaviors of defects are rather
similar as the system goes to amorphous phase. Additionally, it is
natural to ask whether the percolation mechanism is universal for
other amorphization scenarios on surfaces of different geometry. Understanding
the organization of defects can help design particle structures for
multiple applications, such as colloidosomes\citep{dinsmore2002colloidosomes},
photonic crystals\citep{hou2018patterned} or building blocks for
new materials\citep{sacanna2011shape}. Particle anisotropy has also
been shown to play a key role in local unjamming in biological media\citep{grosser2021cell}. 

In this work, we demonstrate that the percolation mechanism occurs
in the original NRS scenario, and for many other kinds of anisotropy
that could be present. We examine: bidispersed mixtures on flat surfaces
as considered by NRS, mixtures of identical elongated particles of
varying aspect ratio $\lambda$, such as ellipsoids, on curved surfaces
of different topology. We also consider a system with long-range interactions,
an anisotropic generalization of the Thomson problem\citep{perez1997influence,bowick2002crystalline,wales2006structure},
whereby mixtures of different charge with ratio $\rho=q_{2}/q_{1}$
are arranged to minimize the electrostatic energy. Henceforth, we
shall unify all these measures of anisotropy by collectively defining
a single parameter $a\in[0,1]$, which depending on the system may
be the bidispersity $\left(R-r\right)/\left(R+r\right)$, shape anisotropy
$(\lambda-1)/(\lambda+1)$ or charge anisotropy $\left(q_{2}-q_{1}\right)/\left(q_{1}+q_{2}\right)$.

\section{Results \& Discussion}

To do so, we generate packings of $N=1000$ particles on flat surfaces,
spherical surfaces and toroidal surfaces with aspect ratio 2, using
a Monte Carlo procedure inspired by the Lubachevsky-Stillinger algorithm\citep{Lubachevsky1990,donev2004jamming}:
particles are initially randomly placed on a large surface, then diffuse
both translationally and rotationally by Brownian motion while the
size of the suface is gradually reduced. After reduction moves, gradient
descent is performed on an objective function that penalizes overlaps.
If overlaps cannot be removed, the algorithm backtracks and reduces
the rate of reduction; the algorithm is halted when the reduction
rate reaches a critical threshold. For bidispersed packings, a fraction
$\chi=\frac{1}{2}$ of particles are inflated. Details of this algorithm
are presented in previous work\citep{burke2015role,mascioli2017defect,xie2019geometry}
and necessary modifications to deal with anisotropic particles are
described in Methods below. For the Thomson problem, all charges are
initially set equal and a minimum is found by conjugate gradient descent;
a fraction $\chi=\frac{1}{2}$ of charges are randomly selected and
increased in magnitude; then the energy is reminimized.

Disclinations are identified by the following procedure: We first
generate a Voronoi diagram that approximates the navigation map\citep{luchnikov1999voronoi,schaller2013set}
from a cloud of points generated to lie on the boundary of the particles;
particles that possess a connected edge in this graph are identified
as neighbors. From the resulting neighbor graph, we find the \emph{subgraph}
of defects, i.e. vertices that have connectivity other than $6$. 

Representative packings as a function of anisotropy $a$ and their
corresponding defect subgraphs are shown in Fig. \ref{fig:Packings-and-associated}A.
Note that the representation of the subgraphs displayed here is designed
to emphasize the topological features; there is no significance to
the spatial position of the nodes. For monodispersed particles, the
packings are crystalline as expected. On the flat surface, a few isolated
defect motifs are typically present because the lattice may be incommensurate
with the periodic boundary conditions. On curved surfaces the scars
of spherical crystallography occur together with a number of dislocations.
While defects are not topologically required on the torus, because
the genus is $1$ and the Euler characteristic is $0$, the higher
curvature present locally deforms the crystal lattice and therefore
tends to promote longer scars and star motifs. As the degree of anisotropy
is increased, the size of the defect motifs increases for all cases,
and, eventually, a system-spanning structure emerges. In these packings,
our focus is the process of elongation of defect networks hence packings
beyond the system-spanning structures are not the interest of this
manuscript.

The packing fraction as a function of anisotropy $a$ for bidispersed
spheres on the flat surface and ellipsoids on the surface of a sphere
or a torus are displayed in Fig. \ref{fig:Packings-and-associated}B.
For the bidispersed mixture, as $a$ increases, the packing fraction
decreases initially due to the introduction of disorder and afterwards
slowly increases because the small particles tend to fill in the gap
between large particles, consistent with the research on spherical
surface\citep{mascioli2017defect}. On the contrary, for ellipsoids,
when $a$ increases, the packing fraction also increases to balance
the additional rotational degree of freedom then decreases due to
exclusion-volume effects, in agreement with previous results\citep{donev2004improving}.

In Fig. \ref{fig:Packings-and-associated}C we show how the fraction
of defects $p$ varies as a function of the relevant anisotropy parameter
$a$, showing that although the detailed variation of $p$ depends
on the particular scenario considered, these have a similar functional
form: As, $a\to0$, $p$ is small and constant. Above a certain value
of $a$, $p$ begins to increase rapidly and eventually saturates.
The value of anisotropy at which defect clusters begin to form, and
the ultimate value of $p$, varies between the scenarios considered;
ellipsoidal packings saturate at a significantly lower $p$ than for
the isotropically shaped particles; increasing the charge ratios in
Thomson problem can achieve a larger $p$ than elongated particles. 

In Fig. \ref{fig:Packings-and-associated}D, we also display the growth
of a spanning structure indicated by the fraction of scar length of
the largest connected defect subgraph among all defects for various
extent of anisotropy $a$ in different systems. The fraction quickly
rises to $1$ as we increase the anisotropy, indicating the percolation
transition occurs with the formation of a globally connected cluster.
Also note that the formation of a system-spanning structure is slower
by varying the shape of particles than by adding bidispersity. We
note that a rescaling of the form $a\to a^{\alpha}$ can be used to
bring the transition points of the different scenarios into alignment
and hence partially collapse the curves in both Fig. \ref{fig:Packings-and-associated}C
and D. The physical significance of these powers remains unclear,
however, and hence understanding the detailed form of these curves
is left to future work.

As the degree of anisotropy and defect motifs increase in size, the
system goes from crystalline phase into the amorphous phase as indicated
by various structural order parameters. The translational order can
be examined by the pair correlation function\citep{giarritta1992statistical,truskett1998structural,donev2005pair,vest2018glassy}
$g(r)=\rho(r)/\rho_{0}$ where $\rho_{0}$ is the overall density
of particles and $\rho(r)$ is the density at distance $r$ from the
centered reference particle. The local orientational order of particle
$i$ can be measured by bond orientational order $\psi_{6}(\overrightarrow{r_{i}})=\sum\exp(i6\theta_{ij})/n$,
where $\theta_{ij}$ is the angle of the bond connecting particle
$i$ and its neighbor $j$ with respect to some local axis and $n$
is the number of nearest neighbors. The bond orientational correlation
function
\[
G_{6}(r=\left|\overrightarrow{r_{i}}-\overrightarrow{r_{j}}\right|)=\left\langle \psi_{6}^{*}(\overrightarrow{r_{i}})\psi_{6}(\overrightarrow{r_{j}})\right\rangle ,
\]
displays the global orientational order\citep{rubinstein1982order,nelson1982order,bagchi1996computer,li2005melting,han2008melting,prestipino2011hexatic,qi2014two,guerra2018freezing}.
Previous work\citep{guerra2018freezing} points out that vector transport
on the curved surface complicates measurement of this quantity in
contrast to flat space where a global reference coordinate system
can be defined, and develops a procedure of selecting proper local
reference axis to calculate $\psi_{6}$ on spherical surfaces, where
the particles are projected onto the faces of the icosahedron whose
vertices are aligned with the position of defects and then the local
reference axis on each face is determined such that they are in the
same direction after the icosahedron is unfolded onto a plane; further
details are presented in Methods below. We use the angle $\theta$
to represent the distance between two particles on spherical surfaces.

\begin{figure}
\begin{centering}
\includegraphics[width=3.5in]{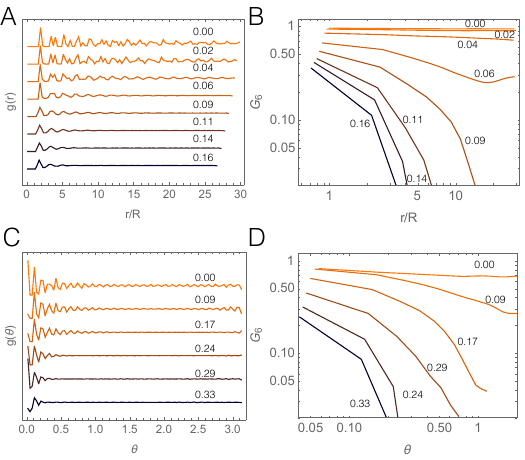}
\par\end{centering}
\caption{\label{fig:orderCorrelation}\textbf{Structural signatures of packings
on flat surfaces and spherical surfaces reveal the transition from
crystalline to amorphous phase with increasing anisotropy.}(A) Pair
correlation function $g(r)$ and (B) bond orientational correlation
function $G_{6}(r)$ for bidispersed packings on flat surfaces. (C)
Pair correlation function $g(\theta)$ and (D) bond orientational
correlation function $G_{6}(\theta)$ for ellipsoidal packings of
different anisotropy on spherical surfaces. Numbers indicate the value
of anisotropy.}
\end{figure}

Fig. \ref{fig:orderCorrelation} displays the evolution of $g$ and
$G_{6}$ with increasing anisotropy for bidispersed packings on flat
surfaces and ellipsoidal particles on spherical surfaces. On flat
surfaces(Fig. \ref{fig:orderCorrelation}A\&B), we recover the NRS
results: the system is in crystalline phase at $b=0$, with both translational
order and orientational order over the whole system. As $b$ goes
above $0.04$, the translational order becomes short range but the
orientational order remains from the algebraical decay of $G_{6}$,
indicating the packing is in the hexatic phase. After that, exponential
decay of $G_{6}$ marks the system entering amorphous phase. For ellipsoidal
packings on spherical surface(Fig. \ref{fig:orderCorrelation}C\&D),
initially the system has long range order without anisotropy. As the
particles are elongated, the long range order is lost with $G_{6}$
turning into algebraical and then exponential decay as the system
transitions into the hexatic phase and then the amorphous phase.

\begin{figure*}
\begin{centering}
\includegraphics[width=5.5in]{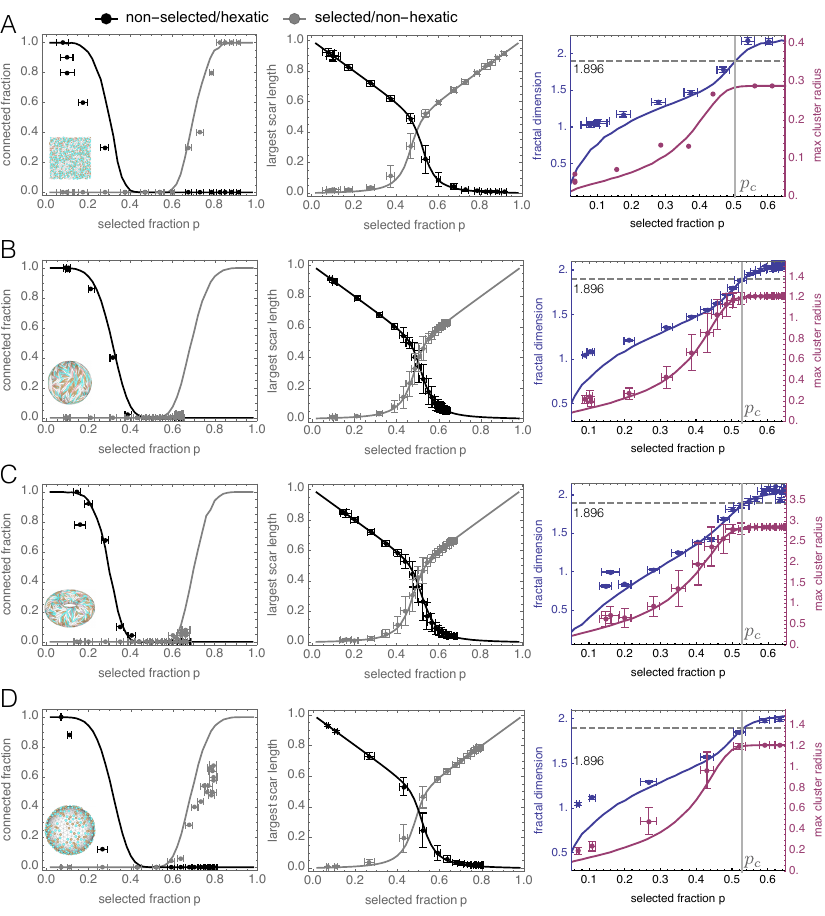}
\par\end{centering}
\caption{\label{fig:DefectStructure}\textbf{Size and structure of defect networks
is well predicted by a percolation model for several different anisotropy
scenarios.} (A) For bidispersed packings on flat surfaces, fraction
of packings with connected spanning structure(left), fraction of particles
in the largest connected component(middle), cluster radius of the
largest connected component and fractal dimensions(right) as a function
of defect fraction $p$. Data points are calculated from an ensemble
of packings as described in the text; solid lines are the predictions
of a percolation model with no fitting parameters. (B)---(D) show
corresponding figures for ellipsoid packings on sphere, ellipsoid
packings on torus and generalized Thomson problem. Error bars indicate
standard deviation computed from our dataset.}
\end{figure*}

We now show that this transition lies in the percolation universality
class, which describes systems with a parameter $p$ that controls
the occupancy of sites or bonds. With increasing $p$, clusters of
connected components arise with increasing size, and above a critical
value $p_{c}$, the mean cluster size diverges for infinite lattices.
The value of $p_{c}$ depends on details of the particular system,
but in the vicinity of $p\to p_{c}$, the cluster size distribution
and structure exhibit universal behavior: For example, the clusters
become fractalline and the cluster radius $R$, 
\begin{equation}
R^{2}=\frac{1}{2}\sum_{i\neq j}\frac{d_{ij}}{n^{2}}\label{eq:clusterRadius}
\end{equation}
where $d_{ij}$ is the distance between pairs of sites $(i,j)$ and
$n$ is the number of sites, scales with the number of sites like
$n\propto R^{D}$ where $D$ is the fractal dimension. In two dimensions
this has a value at $p_{c}$ of $91/48=1.896$ independent of the
structure of the system\citep{stauffer2018introduction,lorenz1993universality,he2002two,han2015shape,zierenberg2017percolation,mitra2019percolation}.
Many disordered systems, both discrete\citep{he2002two,han2015shape,lebovka2006scaling}
and continuous\citep{lee1990monte,lorenz1993universality,quintanilla2000efficient,quintanilla2007asymmetry,yi2002analytical,asikainen2000percolation}
lie in this class, including forest fires, distribution of oil inside
porous rock, the diffusion of atoms and conductivity of electrical
networks\citep{grimmett1999percolation,stauffer2018introduction}.

An important feature of the packing problems considered is that they
involve a finite number of particles, either for reasons of tractability
or because they occur in compact geometries. In finite systems, the
percolation transition becomes second order. For example, the fraction
of simulations $u(p)$ that yield a globally connected cluster as
a function of $p$ is, in an infinite system, the unit step function
$\theta(p-p_{c})$ centered on the percolation point $p_{c}$. At
finite $N$, $u(p)$ become sigmoidal in shape and converges toward
$\theta(p-p_{c})$ as the number of particles is increased. The value
of $p_{c}$ may therefore be extrapolated from a sequence of simulations
of different size\citep{newman2000efficient,he2002two,yi2002analytical,han2015shape}.
An alternative approach is to study the cluster size $R$ as a function
of $p$, which saturates at $p_{R}(N)$ as $R$ approaches the system
size\citep{mascioli2017defect}. The saturation point converges on
$p_{c}$ as $N\to\infty$. These different definitions need not necessarily
coincide in finite systems.

Further, in contrast to the canonical site and bond percolation problems
described above where $p$ is a parameter that can be directly varied---in
these problems $p$ is the fraction of sites or bonds chosen on a
specified lattice---here in packing problems it is the degree of
anisotropy that is varied. The neighbor graphs and defect subgraphs
are not known ahead of time and must be determined from the packing.
We therefore identify $p$ as the fraction of sites that lie in the
defect subgraphs. This identification enables us to make an explicit
comparison between a system that is not manifestly in the percolation
universality class with one that is by construction: In each of scenarios
considered, we create a zero anisotropy packing---which is of course
crystalline---and compute its neighbor graph. We then study the site
percolation problem on this graph, where we randomly select $p$ fraction
of sites iteratively and investigate their structure for every trial. 

We may now examine the growth and structure of the clusters in the
percolation model. The fraction of trials where the selected sites
connect into a spanning structure is counted and showed as gray lines
in left column of Fig. \ref{fig:DefectStructure}. The fraction becomes
nonzero at around $p_{c}=0.5$ on flat surfaces and $p_{c}=0.55$
on spherical or toroidal surfaces, which is in good agreement with
previous literature\citep{he2002two,han2015shape,mascioli2017defect}.We
also compute the fraction of sites in the largest connected component,
displayed as gray lines in the middle column of Fig. \ref{fig:DefectStructure}.
In addition the structures of unselected sites are also computed and
the corresponding results are displayed as black lines, which has
the mirror symmetry compared with those of the selected sites as expected.
The largest cluster radius $R$ and fractal dimension of the selected
sites are shown as a function of $p$ in right column of Fig. \ref{fig:DefectStructure}.
The radius saturates at the percolation transition $p_{c}$ mentioned
above where the value of fractal dimension is consistent with the
universal value of 1.896, marked by the dash lines. 

We further compare the structure of the defect subgraphs with this
percolation model where the non-hexatic defects are recognized as
the selected sites and the hexatic particles are treated as unselected
sites. We compute the fraction of packings whose defects form a globally
spanning structure, the fraction of sites in the largest defect subgraphs,
the largest radius of the defect subgraphs and fractal dimension.
The corresponding quantities are overlaid onto the curves of percolation
model as dots in Fig. \ref{fig:DefectStructure}, described quite
well by the model with no fitting parameters. The deviation of those
dots from the curves is likely due to the finite-size effect. To plot
those curves from the percolation model, we selected fraction $p$
of particles thousands of times on the same lattice. However we can
only generate a finite number of packings. Another possible reason
is the existence of rattlers\citep{donev2004jamming}, i.e. particles
that don't contribute to the connected networks. Therefore these rattlers
can slightly reduce the fraction of packings that have a system-spanning
structure. Nonetheless the agreement, particularly around the percolation
point, is very good indeed, showing that the growth and structure
of the clusters are well predicted by percolation theory. Note that
for ellipsoid packings showed in Fig. \ref{fig:DefectStructure}B
and C, the fraction of defects, represented by selected fraction,
just exceeds the percolation threshold since the defect ratios saturate
at around that value, as showed in Fig. \ref{fig:Packings-and-associated}C.

\section{Conclusions}

In this work, we have considered the emergence, elongation and global
connection of defect structures as a function of different kinds of
anisotropy. Besides bidispersity on flat and curved surfaces that
link our work to the well-explored KTHNY transition, we demonstrate
that elongating particles, or changing the nature of the interaction
have a similar effect: anisotropy induces dislocations that cause
the system to successively lose translational and orientational order,
during which process the newly generated dislocations gradually form
a globally connected cluster. We have further shown that structural
features of the clusters, e.g. fractal dimension, are well predicted
by the percolation model.

Our results suggest an apparent universality in that the defect structures
that emerge when adding anisotropy to a crystalline system appear
to be independent of the source of anisotropy. Bidispersity on the
flat surface or on the surface of a sphere yield similar results to
elongating particles or soft long-range interactions, or to packings
on the surface of a torus. Intriguingly, our results for ellipsoidal
particles show that the defect fraction only just exceeds the percolation
threshold: we speculate that there may exist kinds of anisotropy that
do not yield percolating defects and hence suppress the amorphous
phase. One possible strategy to do so is to consider mixtures of particles
that together form a tessellating structure; e.g. octagons and suitably
sized squares, girih tiles\citep{lu2007decagonal}. Continuously deforming
from uniform spheres towards such special configurations might eliminate
the percolation effect. We suspect that other strategies might exist,
perhaps involving non-convex particles for example as recent paper
shows they can change the geometrical percolation threshold\citep{lin2018geometrical},
and this newfound connection between particle shape and defect structure
should open new avenues for tunability in the mechanics of particulate
media.

\section{Methods}

\subsection*{Packing algorithm for ellipsoidal particles}

Our algorithm to pack spherical particles is as described above and
in previous work\citep{burke2015role,mascioli2017defect,xie2019geometry}.
The extension to ellipsoidal particles involves modifications to overlap
detection and diffusion as follows: The range parameter $\sigma_{ij}$
is given by,

\[
\sigma_{ij}=2b/\sqrt{1-\frac{\chi}{2}(\frac{\hat{d_{ij}}\cdot\hat{u_{i}}+\hat{d_{ij}}\cdot\hat{u}_{j}}{1+\chi(\hat{u_{i}}\cdot\hat{u_{j}})}+\frac{\hat{d_{ij}}\cdot\hat{u_{i}}-\hat{d_{ij}}\cdot\hat{u_{j}}}{1-\chi(\hat{u_{i}}\cdot\hat{u_{j}})})},
\]
where $a$ and $b$ are the half lengths of the major and minor axes,
$\chi=(a^{2}-b^{2})/(a^{2}+b^{2})$, $\hat{u}$ is a unit vector describing
the overall orientation of the particle and $\hat{d_{ij}}$ is the
unit vector pointing from one center to the other. If the center-to-center
distance $d_{ij}$ is smaller than the range parameter $\sigma_{ij}$,
there is overlap. This criterion has been successfully implemented
in other work\citep{padilla1997isotropic,anquetil2012competing,zeravcic2009excitations,debenedictis2015competition}.
If two ellipsoidal particles overlap, we exert the Gaussian model
potential\citep{berne1972gaussian}
\[
V(u_{i},u_{j},d_{ij})=\epsilon_{0}\left[1-\chi^{2}(\hat{u_{i}}\cdot\hat{u_{j}})^{2}\right]^{-1/2}\exp\left(-d_{ij}^{2}/\sigma_{ij}^{2}\right),
\]
where $\epsilon_{0}$ is the strength parameter, to remove the overlaps
by gradient descent.

Diffusion of particles is another feature of our algorithm. Spheres
can diffuse simply by Langevin equation 
\[
\mathbf{x}_{i}^{'}(t+\delta t)=\mathbf{x}_{i}(t)+\mathbf{\eta}_{i}\sqrt{2D\delta t},
\]
where $\eta_{i}$ is a random step drawn from Gaussian distribution,
$D$ is the diffusion constant such that $\sqrt{2D\delta t}$ determines
the variance of the displacement for a timestep $\delta t$. For ellipsoids,
we must also account for rotations. First we rotate the director of
an ellipsoid by 
\[
\delta\theta(\delta t)=\eta_{\theta}\sqrt{2D_{\theta}\delta t}.
\]
Then in the local coordination system $\widetilde{x}$ and $\widetilde{y}$
along the major and minor axes, it is displaced by 
\[
\delta r(\delta t)=\eta_{a}\sqrt{2D_{a}\delta t}\widetilde{x}+\mathbf{\eta}_{b}\sqrt{2D_{b}\delta t}\widetilde{y}.
\]
Finally we transform this local displacement into the global coordinate
system by multiplying the rotation matrix\citep{han2006brownian}.

\subsection*{Orientational correlation on spherical surfaces}

Here we describe how to find the icosahedron that align with the defects
following the method previously reported\citep{guerra2018freezing}.
A non-trivial icosahedrally symmetric function can be defined as 
\[
h_{6}(\hat{x})=Y_{6,0}(\hat{x})+\sqrt{7/11}(Y_{6,-5}(\hat{x})-Y_{6,5}(\hat{x})),
\]
where $Y$ is the spherical harmonics. The positions of local max
values align with the vertices of an icosahedron. The defects on the
spherical surface can be rotated by the rotational matrix 
\[
R_{\omega}(\theta,\phi)=\left(\begin{array}{ccc}
\cos\theta & 0 & \sin\theta\\
0 & 1 & 0\\
-\sin\theta & 0 & \theta
\end{array}\right)\left(\begin{array}{ccc}
\cos\phi & -\sin\phi & 0\\
\sin\phi & \cos\phi & 0\\
0 & 0 & 1
\end{array}\right).
\]
Then $(\theta,\phi)$ is computed by minimizing $\sum_{i}h_{6}(R_{\omega}(\theta,\phi)\cdot\overrightarrow{r_{i}})$
where $\overrightarrow{r_{i}}$ represents coordinates of defects.
The inverse rotation gives the icosahedron that align with the defects.

\section*{Conflicts of interest}

There are no conflicts to declare.

\section*{Acknowledgments}

This material is based upon work supported by the National Science
Foundation under Grant No. DMR-1654283. One of the authors (TJA) thanks
the Kavli Institute for Theoretical Physics ACTIVE'20 program which
was supported in part by NSF Grant No. PHY-1748958, NIH Grant No.
R25GM067110, and the Gordon and Betty Moore Foundation Grant No. 2919.02

\bibliographystyle{rsc}
\bibliography{preprint}

\end{document}